\documentstyle[12pt,epsf,epsfig]{article}
\def\lsim{\mathrel{\rlap {\raise.5ex\hbox{$ < $}}
{\lower.5ex\hbox{$\sim$}}}}
\def\gsim{\mathrel{\rlap {\raise.5ex\hbox{$ > $}}
{\lower.5ex\hbox{$\sim$}}}}

\newcommand{\pr}{\paragraph{}}
\newcommand{\be}{\begin{equation}}
\newcommand{\ee}{\end{equation}}
\newcommand{\bea}{\begin{eqnarray}}
\newcommand{\nn}{\nonumber}
\newcommand{\eea}{\end{eqnarray}}
\newcommand{\nd}[1]{/\hspace{-0.6em} #1}
\newcommand{\nk}{\noindent}
\def\gappeq{\mathrel{\rlap {\raise.5ex\hbox{$>$}}
{\lower.5ex\hbox{$\sim$}}}}
 
\def\lappeq{\mathrel{\rlap{\raise.5ex\hbox{$<$}}
{\lower.5ex\hbox{$\sim$}}}}
 
\begin{document}
 
\begin{titlepage}
\vspace*{2cm} 
\begin{flushright}
CERN-TH/96-189 \\
LBNL-39049 \\
ACT-10/96 \\
CTP-TAMU-31/96 \\
hep-ph/9607434
\end{flushright}

\begin{centering}
\vspace{.1in}
{\large {\bf 

CPT and Superstring$^*$
}} \\
\vspace{.2in}
{\bf John Ellis$^a$, N.E. Mavromatos$^{b}$ and D.V. Nanopoulos$^c$}\\
\vspace{.1in}
\vspace{.03in}
\vspace{.1in}
{\bf Abstract} \\
\vspace{.05in}
\end{centering}
{We discuss the possibility that CPT violation may appear as a
consequence of microscopic decoherence due to quantum-gravity effects,
that we describe using a density-matrix formalism motivated by our
studies of non-critical string theory. The maximum possible
order of magnitude of such decohering CPT-violating
effects is not far from the sensitivity of
present experiments on the neutral kaon system, and we review a 
simple parametrization for them. We also review a recent data analysis
carried out together with the CPLEAR collaboration, which bounds any
such decohering CPT-violating parameters to be $\lappeq 10^{-19}$ GeV.}
\vspace{0.2in}
%
\pr

\vspace{0.01in}
\begin{flushleft}
$^a$ Lawrence Berkeley National Laboratory, University of
California, Berkeley, California 94720, \\
on leave from Theory Division, CERN, CH-1211, Geneva, Switzerland.  \\
$^b$ P.P.A.R.C. Advanced Fellow,
University of Oxford, Dept. of Physics
(Theoretical Physics),
1 Keble Road, Oxford OX1 3NP, United Kingdom.   \\
$^{c}$ Center for
Theoretical Physics, Dept. of Physics,
Texas A \& M University, College Station, TX 77843-4242, USA
and Astroparticle Physics Group, Houston
Advanced Research Center (HARC), The Mitchell Campus,
Woodlands, TX 77381, USA. \\
\end{flushleft}

\vspace{0.1in}

\begin{flushleft}
$^*$ Invited talk by J.E. at the {\it Workshop on K Physics},
Orsay, France, May 30 - June 4, 1996; \\
This work was supported in part by the Director, Office of Energy
Research, Office of Basic Energy Science of the U.S. Department of
Energy, under Contract DE-AC03-76SF00098. \\
\end{flushleft}

\vspace{0.1in}
\begin{flushleft}
July 1996
\end{flushleft}
\end{titlepage}

\newpage
\section{CPT Invariance}

\pr
This is a fundamental theorem of Quantum
Field Theory, which follows from the basic requirements of
locality, Lorentz invariance and unitarity~\cite{lud}. Among the 
consequences of the CPT theorem are equal masses for particles
and their antiparticles, equal lifetimes, equal and opposite
electric charges, and equal magnetic moments. None of these
properties need hold in non-relativistic quantum mechanics, which 
need not obey all the constraints of local quantum field theory.
However,  what interests us is the possibility
that CPT invariance may be violated as a consequence of
microscopic decoherence 
in the context of quantum gravity in general, and specifically
in the context of string theory.

\pr
A number of experimental upper limits on violations of the CPT theorem
can be found in the Particle Data Book~\cite{pdg}. They include 
the following
measurements on electrons and positrons:
\begin{equation}
|{\Delta m_e \over m_e}| < 4 \times 10^{-8},
|{\Sigma Q_e \over Q_e}| < 4 \times 10^{-8},
{\Delta ( g - 2)_e \over g_e} = (-0.5 \pm 2.1) \times 10^{-12},
\label{cpte}
\end{equation}
the following test with muons:
\begin{equation}
{\Delta ( g - 2 )_{\mu} \over g_{\mu}} = (-2.6 \pm 1.6) \times 10^{-8}
\label{cptmu}
\end{equation}
and the following test with protons:
\begin{equation}
{\Delta m_p \over m_p} = ( 2 \pm 4 ) \times 10^{-8}.
\label{cptp}
\end{equation}
These are all very impressive limits, but they all pale in
precision compared with the bound from the neutral kaon system:
\begin{equation}
{\Delta m_K \over m_K} < 9 \times 10^{-19}
\label{cptk}
\end{equation}
obtained by the CPLEAR collaboration
\cite{dmk}.

\pr 
In view of the impeccable credentials of the CPT theorem, and the
tremendous accuracy with which it has been verified, why would any
theorist challenge it, and why should any experimentalist want to try any
harder than in (\ref{cpte},\ref{cptmu},\ref{cptp},\ref{cptk}) 
above? One possible
motivation is provided by the apparently unrelated theoretical problem
discussed in the next section. 

\section{Do Topological Space-Time Fluctuations Destroy
Quantum Coherence?}

\pr
It is known~\cite{hawkbek} 
that a macroscopic four-dimensional black hole
has non-trivial entropy $S$ related to the area $A$ of its event horizon,
and hence to its mass $M$ in the case of a black hole with no 
additional quantum numbers such as electric charge $Q$ or spin
$J$~\footnote{Here and subsequently, we use natural units in which
the Planck mass $M_P = 1$}:
\begin{equation}
S\,=\,{1 \over 4} A \, = M^2
\label{entrbh}
\end{equation}
A macroscopic black hole also has an effective temperature $T$:
\begin{equation}
T \, = \, {1 \over 8 \pi M}
\label{tempbh}
\end{equation}
as manifested, for example, by its Hawking radiation.
The facts that the entropy (\ref{entrbh}) and temperature of
the macroscopic black hole are non-zero tell us immediately that it
{\it must} be described by a mixed quantum-mechanical state.
At first sight, this is surprising, because we could certainly imagine 
having made our black hole by colliding particles in a pure initial
state, and conventional quantum mechanics and quantum field theory
forbid pure states from evolving into mixed states.

\pr
The basic intuition behind the appearance of a mixed state is
the following: imagine that a pure state with two 
components $|A,B>$ is prepared near the event horizon of a
macroscopic black hole, and that one of the components, $|B>$ say,
falls inside the horizon. Conventional semiclassical quantum
gravity, as embodied in Hawking's original calculation of the
radiation that bears his name, 
would suggest that all information $I$ about the 
component  $|B>$ is lost to an exterior observer, including for 
example information about its quantum-mechanical phase, and that
all one can therefore observe is a mixed external component with a 
density matrix
\begin{equation}
\rho_A \, = \Sigma_I |A>_I {}_I <A|
\label{mixed}
\end{equation}
suggesting the ``forbidden" evolution of a pure state into a mixed
state.

\pr
This proposal is already controversial with many quantum field 
theorists, but 
what makes them really see red~\cite{gross} is the further suggestion 
of Hawking~\cite{hawk2} 
that such evolution from pure into mixed states
might also occur at the microscopic level. Here the hypothesis is
that information about particle wave functions, etc., might be
lost across microscopic event horizons associated with
topologically non-trivial fluctuations in the space-time background
taking place on the Planck distance scale $L_P \simeq 10^{-33}$ cm
within the Planck time scale $t_P \simeq 10^{-43}$ seconds, called
generically space-time foam. The intuition behind any such
proposal is an analogy with the conventional quantum mechanics of
open systems coupled to undetected degrees of freedom, 
represented in this case by physics at the Planck scale $L_P, t_P$.
Here, however, the suggestion is that this might be an intrinsic
and fundamental limitation of laboratory physics, rather than 
an artefact of technological limitations or laziness in not
measuring the state of some communicating reservoir. 

\pr 
Any such fundamental loss of information, and hence transition from
pure to mixed states at the microscopic level, would entail a modification
of conventional quantum field theory and quantum mechanics, which is
repugnant to many theorists enamoured of the great beauty of these
theories. 

\pr
Concretely, Hawking has proposed~\cite{hawk2} that one should abandon the
conventional $S$-matrix description of asymptotic particle
scattering, and replace it by a density matrix description, in
which scattering occurs via a superscattering operator $\nd{S}$:
\begin{equation}
{\rho_{out}}^A_B \, = \, \nd{S}^{AD}_{BC} \, {\rho_{in}}^C_D
\label{dollar}
\end{equation}
where, unlike in conventional field theory, $\nd{S}$ does not 
factorize into a product of matrix elements of $S$ and its 
hermitian conjugate:
\begin{equation}
{\nd{S}}^{AD}_{BC} \, \ne \, S^A_C \, {S^+}^D_B
\label{nofact}
\end{equation}
One way of thinking about this proposal may be that, whereas $\nd{S}$
factorizes in any fixed space-time background, this would no
longer be the case if one sums over many such backgrounds, as may be
appropriate to take into account fluctuations in the space-time foam.

\pr
If the proposal (\ref{dollar},\ref{nofact}) is correct, a corollary
would be that the Liouville equation that describes the time evolution
of a quantum system must also be modified. When integrated over all
time, the usual Liouville equation $\partial _t \rho = i [\rho, H]$
becomes the normal $S$-matrix
description of asymptotic scattering. In order to avoid this so as to 
accommodate a lack of factorization
(\ref{nofact}), we need an extra term~\cite{ehns}:
\begin{equation}
\partial _t \rho = i [\rho, H] + \nd{\delta H} \rho
\label{modliou}
\end{equation}
Just such a modification of the quantum Liouville equation is
characteristic of open quantum-mechanical systems. Here, the
`openness' would be due to the coupling of the observable system
to unseen (unseeable?) modes of the theory within the 
microscopic event horizons believed to infest space time. Any
modification (\ref{modliou}) would necessarily cause pure states
to evolve into mixed states, with the collapse of off-diagonal entries
in the density matrix. One can also show that symmetries no longer
correspond to conservation laws, in general~\cite{ehns}. 

\pr
Clearly, any modification of the type (\ref{modliou}) should
respect probability conservation, which means that Tr$\rho$
should be time-independent, and it should also conserve energy, 
at least to a very good approximation. Shortly after the proposal
(\ref{modliou}) was made, the concern was expressed 
\cite{BPS} that energy non-conservation
might be the Achilles heel of this idea. However, we have shown
that
energy is conserved, in the sense that 
\begin{equation}
\partial _t < E > \, = \, 0 \, : \, <E> \, \equiv \hbox{Tr}(E \rho)
\label{econs}
\end{equation}
in our non-critical
string approach~\cite{emn}, as a consequence of the renormalizability of the
two-dimensional field theory on the string world sheet.
Moreover, Unruh and Wald have recently given general arguments why
energy conservation need not be an essential difficulty~\cite{unruh}.
\pr
In order to gain some intuition how solutions to a modified
quantum Liouville equation of the type (\ref{modliou}) behave, it is
instructive to consider~\cite{ehns} the simplest possible example of
a two-state system with two energy levels $E \pm \Delta E/2$:
\begin{equation}
H \, = \, E \, + \, \Delta E \sigma_z
\label{twostate}
\end{equation}
whose normal quantum-mechanical evolution is unitary: 
\be
\rho (t) =\frac{1}{2}\left(\begin{array}{c} 1 \qquad
e^{-i\Delta E t} \\ e^{i\Delta E t} \qquad 1 \end{array}\right)
\label{unitary}
\end{equation}
as seen in Fig.~1(a).

\begin{figure}
\hglue2.5cm
\epsfig{figure=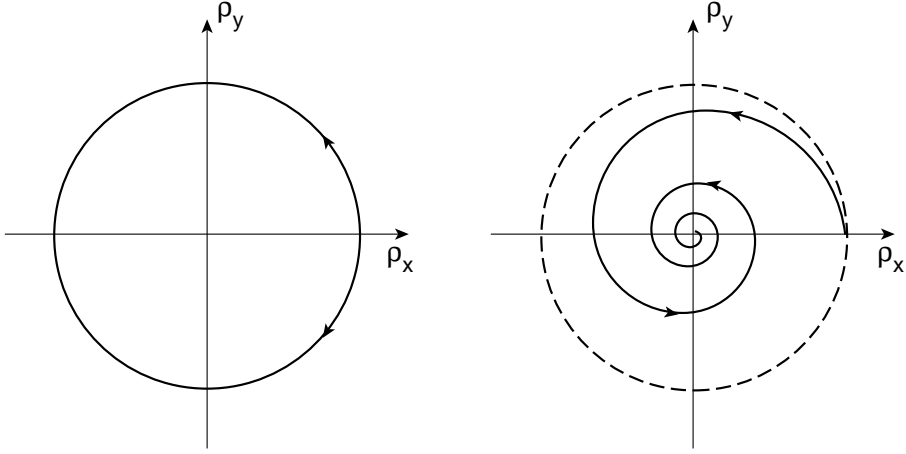,width=12cm}
\caption[]{The unitary evolution of a simple two-dimensional
quantum-mechanical system is contrasted
with the spiralling behaviour induced by the parameters
$\alpha, \beta, \gamma$, representing evolution towards a
completely mixed diagonal density matrix \cite{ehns}.}
\end{figure}

\pr
In the open quantum-mechanical formalism (\ref{modliou}), 
it is convenient to parametrize $\rho$
in the four-dimensional
Pauli $\sigma$-matrix basis $\sigma_{\alpha}$ 
for $2 \times 2$ hermitian matrices
$({\bf 1}, \sigma _x, \sigma_y, \sigma _z)$:
$\rho = \rho_{\alpha} \sigma_{\alpha}$. In this formalism,
one can regard $\nd{\delta H}$ as a symmetric $4 \times 4$
matrix $\nd{h}_{\alpha \beta}$. Probability
conservation then requires
\begin{equation}
\nd{h}_{0 \beta} \, = \, 0 \, = \, \nd{h}_{\alpha 0}
\label{2pcons}
\end{equation}
and energy conservation requires
\begin{equation}
\nd{h}_{3 \beta} \, = \, 0 \, = \, \nd{h}_{\alpha 3}
\label{2econs}
\end{equation}
so that its general form is
\be
\nd{ h}_{\alpha \beta} =\left(\begin{array}{c}
0 \qquad 0 \qquad 0 \qquad 0 \\
0 \qquad -\alpha \qquad -\beta \qquad 0 \\
0 \qquad -\beta \qquad -\gamma \qquad 0 \\
0 \qquad 0 \qquad 0 \qquad 0 \end{array}\right)
\label{hslash}
\ee
where the three free parameters must obey the positivity conditions
\begin{equation}
\alpha , \, \gamma \, > \, 0, \qquad \alpha \gamma \, > \, \beta^2
\label{positiv}
\end{equation}
It is easy to verify that the corresponding evolution of the density 
matrix is non-unitary:
with the addition of the terms (\ref{hslash}), $\rho$ evolves as
\be
\rho (t) = \frac{1}{2} \left(\begin{array}{c}
1 \qquad e^{-(\alpha + \gamma)t/2}e^{-i\Delta E t} \\
e^{-(\alpha + \gamma)t/2}e^{i\Delta E t} \qquad 1 \end{array}\right)
\label{nonunit}
\ee
which spirals into the origin as in Fig.~1(b).
approaching the completely mixed form
\be
\rho (\infty) =\frac{1}{2} \left(\begin{array}{c}
1 \qquad 0 \\
0 \qquad 1 \end{array}\right)
\label{mixedf}
\end{equation}
at large times, with a decay time scale $\tau = 2/(\alpha + \gamma)$.
This toy example turns out to be directly applicable to the $2 \times 2$
neutral kaon system which we discuss later.

\section{Decoherence and CPT Violation}

\pr 
The non-unitary evolution in the simple two-state example above
manifests an arrow of time: the system spirals in, not out. 
Everyday experience tells us that an arrow of
time is present macroscopically: our bit (at least) of the
Universe is expanding, and we are all of us getting older. 
On the other hand, no such arrow
of time is visible in our accepted fundamental laws of physics: Quantum
Field Theory is invariant under CPT, and time $t$ is just a coordinate in
General Relativity - the motion of the Earth around its solar orbit could
be reversed with no apparent problem. On the other hand, an arrow of time
appears in thermodynamics via the second law, which states that entropy
increases monotonically. Is it possible that this has a microscopic
origin?

\pr
It has been pointed out by Wald and Page~\cite{wald} that a 
microscopic arrow of time must appear if pure
states evolve into mixed states as suggested above, in the sense that
the strong form of the CPT theorem must be violated. Suppose there is
some CPT symmetry transformation $\Theta$ which maps initial-state
density matrices into final-state density matrices:
\begin{equation}
\rho'_{out} \, = \, \Theta \rho_{in}
\label{inout}
\end{equation}
and correspondingly
\begin{equation}
\rho'_{in} \, = \, \Theta \rho_{out}
\label{outin}
\end{equation}
where 
\begin{equation}
\rho_{out} \, = \, \nd{S} \rho_{in}, \, 
\rho'_{out} \, = \, \nd{S} \rho'_{in}
\label{page}
\end{equation}
It is easy to deduce from these equations that $\nd{S}$
must have an inverse:
\begin{equation}
\nd{S}^{-1} \, = \, \Theta^{-1} \nd{S} \Theta^{-1}
\label{inverse}
\end{equation}
which cannot be true if pure states evolve into mixed states,
entropy increases monotonically and the density matrix
collapses as in the simple two-state example given above.

\pr
Although there are many people in the Quantum Gravity
community who suspect that some modification of Quantum Mechanics
may be necessary so as to incorporate decoherence associated with
black holes, there is disagreement whether this is necessarily
accompanied by CPT violation. This division of opinion is exemplified
by the viewpoints of Hawking and Penrose in \cite{hpbook}: Hawking
is very reluctant to give up CPT, whereas Penrose accepts it as a
likelihood. The formalism we have developed definitely points in
the latter direction, as we see explicitly in connection with the
neutral kaon system in section 5.

\section{CPT Violation and Decoherence in String Theory}

\pr
We have already reminded ouselves that CPT invariance is a
fundamental theorem of string theory, following from its
locality, Lorentz invariance and unitarity. It is clear why
one should re-examine the theorem's validity in string theory.
Strings are described by a local field theory on the two-dimensional 
world sheet, but are not local in four-dimensional space time, being 
extended objects with sizes of the order of the Planck length.
Moreover, although Lorentz invariance is a property of classical
string vacua, corresponding to conformal field theories on the world 
sheet known as critical string theories, this is no longer guaranteed
when one ventures `off shell' into non-critical string 
theories~\cite{emn}.
Various authors have studied conditions under which CPT violation
could indeed occur in string theory. It has been shown that they are not 
met by closed strings in conventional 
fixed flat space-time backgrounds~\cite{wittson}. One
possibility is that CPT might be violated spontaneously in certain
string backgrounds~\cite{kostel}, but 
this would not be correlated with a possible
breakdown in quantum coherence in the way described in the previous
section. 

\pr
It seems to us that such a breakdown is possible
when one treats quantum fluctuations in space time using 
non-critical string theory~\cite{emn}. 
We argue that this leads generically to
decoherence and CPT violation through parameters analogous to the
$\alpha, \beta, \gamma$ in the simple two-state system 
described above. These parameters are theoretically distinct 
and experimentally distinguishable from the $K {\bar K}$
mass and lifetime difference parameters $\delta m, \delta \Gamma$
and the decay amplitude difference $\delta A_{2\pi}$ discussed by
other authors~\cite{others}.

\pr
Before we develop our point of view, we recall that it
is very controversial, and is disregarded by most string 
theorists~\cite{polch}.
They usually think that the study of black holes in string theory
will reveal how quantum-mechanical purity can be maintained. 
This belief is often based on indications that the
black-hole entropy (\ref{entrbh}) can be understood as the
number of distinct string states, as we suggested in 1992 on
the basis of our studies of two-dimensional string black 
holes~\cite{emnstates}.
However, the possibility that pure states might evolve into
mixed states does not shock people who analyze 
the black-hole information problem from a quantum-gravity 
point of view~\cite{hpbook}. The reason we advocate this point of view is
that the types of measurements needed to identify the black-hole
state and all its quantum numbers, such as generalized Aharonov-Bohm
phase measurements using highly-excited string 
states~\cite{emnmeas,banksd}, are not
carried out in practice and are very probably impossible in principle. 
We believe that the full enormity of this problem will become
apparent to other string theorists when they get as far as treating
the back reaction of low-energy particles on black-hole backgrounds,
and discussing quantum fluctuations in the background space time, as
we have been attempting for some time using the two-dimensional black
hole as a representative example. 

\pr
Our string approach starts from the remark that laboratory
experiments are all carried out using light particles such as
$K, n, \pi, \nu$, which consist of lowest-level string degrees of
freedom. Laboratory experiments do not use the full infinite set of
higher-level degrees of freedom that complete the string spectrum.
This would not be a problem if the higher-level states decoupled from 
the lowest-level states, as they do in a fixed, flat space-time  
background. However, it is known that the higher-level states are
indeed coupled to the lowest states in a generic curved background
such as the two-dimensional black 
hole~\cite{chaulykk}, as a result of the same
infinite set of symmetries that provide black-hole quantum numbers
and label the distinct black-hole states ($W_{\infty}$ in the 
two-dimensional case~\cite{emn}).

\pr
The effective low-energy theory is obtained by integrating over the
unseen higher-level states:
\begin{equation}
\tilde \rho (light, t) \, = \, \int d(higher)\rho (light, higher)
\label{intout}
\end{equation}
where the $higher$ states play a role analogous to those of the
unseen states $|B>_I$ inside the black-hole horizon in (\ref{mixed}).
The integration over $higher$ in (\ref{intout})
ensures that the reduced density matrix $\tilde \rho$ is mixed in
general, even if the full $\rho (light, higher)$ is pure. We have
argued that $\tilde \rho$ obeys a modified quantum Liouville
equation (\ref{modliou}) of the form~\cite{emn}
\begin{equation}
\partial _t {\tilde \rho} = i [{\tilde \rho}, H] + \nd{\delta H} \rho 
\qquad : \qquad 
\nd{\delta H} = -i \Sigma_{i,j} \beta^i G_{ij}[\, \, , q^j]
\label{stringmodliou}
\end{equation}
where $H$ is the usual light-particle Hamiltonian, the indices
$(i,j)$ label all possible microscopically-distinct string
background states with coordinate parameters $q^i$, and $G_{ij}$
is a metric in the space of such possible backgrounds~\cite{zam}. We argue
that these are not conformally invariant once one integrates out
the $higher$ degrees of freedom, and the $\beta^i$ are the
corresponding renormalization functions. 
These are non-trivial
to the extent that back reaction of the light particles on the
background metric cannot be neglected. 
Equations of the form (\ref{stringmodliou})
are quite generic in the context of non-critical string 
theories~\cite{emn,kogan}.
\pr
The maximum effect that we can imagine is of order
\begin{equation}
\nd{\delta H} \simeq H^2 / M_P
\label{order}
\end{equation}
which would be around $10^{-19} ... 10^{-20}$ GeV for the neutral 
kaon system. A contribution to the evolution rate equation
(\ref{modliou}) of this order of magnitude would arise if there
were some Planck-scale interaction contributing an amplitude
$A \simeq 1/M_P^2$ and hence a rate $R \simeq 1/M_P^4$, to be
multiplied by a density $n \simeq L_P^{-3} \simeq M_P^3$, yielding the
overall factor of $ \simeq 1/M_P$ shown in 
(\ref{order})~\cite{mohanty,amelino}. A similar
estimate was found in a pilot study of a scalar field in a 
four-dimensional black-hole background~\cite{elizabeth}.

\pr
The origin of the time-dependence on the left-hand side in
(\ref{intout}), which is crucial to the derivation of 
(\ref{stringmodliou}), merits further discussion
at a more technical level. In order to describe
microscopic quantum fluctuations in the space-time background, 
we need to go `off shell', so as to be able to interpolate
between different conformal (critical) string backgrounds, which
is necessary for the description of transitions between them.
Non-conformal backgrounds necessarily lead to divergences which
must be regularized by introducing a cutoff or renormalization scale.
{\it We identify} this with the Liouville field $\phi$, a scalar
field that sets the scale for the world-sheet metric~\cite{emn}:
\begin{equation}
\gamma_{\alpha \beta} \, = \, e^{\phi} \, {\tilde \gamma}_{\alpha \beta}
\label{liouville}
\end{equation}
where $\tilde \gamma$ is a reference metric. In a conformal background,
the dynamics of the Liouville field $\phi$ is trivial, and it decouples
from the rest of the theory. This is no longer true in a non-conformal
background, and one can show that $\phi$ has $negative$ metric. 
{\it We therefore identify} $\phi$ with the target time variable $t$. 
Motion in the space of non-conformal field theories is governed by the 
renormalization-group flow in the $t$ variable, which is controlled
classically by the Zamolodchikov function and the $\beta^i$~\cite{zam}, 
but is subject to quantum fluctuations induced by higher-genus
configurations of the world sheet~\cite{dbrane}. The non-conformal 
divergences induce $t$ dependences that cannot be subsumed in the usual
Hamiltonian evolution of the effective theory of the light string
degrees of freedom, and the extra terms take the form shown in
(\ref{stringmodliou}). We have exhibited explicit terms of this type
associated with transitions between two-dimensional
black holes of different masses, and with the creation and destruction 
of two-dimensional black holes~\cite{emn}.

\pr
One note of clarification is perhaps useful: we are not arguing that
Quantum Field Theory should be abandoned. In our approach, it applies
with all its normal rules to physics on the world sheet. Our point is
that problems may arise in the elevation of quantum physics to target
space, if the latter has singularities and/or a classical event horizon.
The `ugliness' of the $\tilde \rho(light,t)$
density matrix formalism we propose in (\ref{intout}) is not
intrinsic: as we remarked earlier, the full $\rho(light,higher)$ may
well be pure.

\pr
Even if you do not follow the arguments leading to the string version
(\ref{stringmodliou}) of the modified Liouville equation (\ref{modliou}),
the latter still provides an interesting phenomenological framework
in which one can parametrize possible decoherence and CPT-violating
effects with a view to the experimental tests in the neutral kaon
system, which are reviewed in the next section.

\section{Testing Quantum Mechanics and CPT in the Neutral Kaon System}

\pr
This audience does not need convincing that the neutral
kaon system has an enviable track record as a probe of fundamental
physics, ranging from P violation (the $\tau$-$\theta$ puzzle) and
CP violation to the motivation for charm coming from the absence of
strangeness-changing transitions. It is also known to provide very 
elegant tests of quantum mechanics, and provides the most stringent
available test of CPT at the microscopic level, as we saw in section 1.
How can the formalism of decoherence and 
related CPT violation developed
above be applied to the neutral kaon system, and how sensitively
can we test them? 

\pr
The quantum-mechanical Hamiltonian for neutral kaons can
be written as 
\be
  H = \left( \begin{array}{c}
 (M + \frac{1}{2}\delta M) - \frac{1}{2}i(\Gamma + \frac{1}{2}
 \delta \Gamma)
   \qquad  \qquad
   M_{12}^{*} - \frac{1}{2}i\Gamma _{12} \\
           M_{12}  - \frac{1}{2}i\Gamma _{12}
    \qquad  \qquad
    (M - \frac{1}{2}
    \delta M)-\frac{1}{2}i(\Gamma
    - \frac{1}{2}
    \delta \Gamma ) \end{array}\right)
\label{qmmatrix}
\ee
in the $K,{\bar K}$ basis, where $M$ and $\Gamma$ are
the common mass and width of the $K$ and $\bar K$, and
$M_{12}, \Gamma_{12}$ are the complex off-diagonal mixing
parameters. In writing (\ref{qmmatrix}), we have allowed
for quantum-mechanical parameters $\delta M$ and $\delta \Gamma $
that violate $CPT$~\cite{dalitz} but preserve coherence. 
In the density matrix formalism, the time
evolution is given by
\begin{equation}
\partial _t \rho \, = \, -i(H \rho - \rho H^+)
\label{qmevol}
\end{equation}
which preserves the purity of the intial state: it is easy
to see that the asymptotic form of the density matrix in the
$K_{1,2} = (K \pm {\bar K})/\sqrt(2)$ basis is
\begin{equation}
\rho \simeq e^{-\Gamma _Lt}
 \left( \begin{array}{c}
 1   \qquad  \epsilon^* + \delta^* \\
 \epsilon + \delta \qquad |\epsilon + \delta |^2 \end{array}\right)
\qquad : \qquad
  \epsilon =\frac{\frac{1}{2}i Im \Gamma _{12} - Im M_{12}}
{\frac{1}{2} \Delta \Gamma - i\Delta M }, \qquad
\delta    \simeq -\frac{1}{2}
\frac{ \frac{1}{2}\delta \Gamma - i\delta M}
{\frac{1}{2} \Delta \Gamma - i\Delta M}
\label{qmkl}
\end{equation}
where $ \Delta M$ = $M_L$ - $M_S$ is positive and $ \Delta  \Gamma   =
  \Gamma _L -   \Gamma _S$ is negative.
\pr
In our approach, the quantum-mechanical evolution equation
(\ref{qmevol}) is modified to become
\begin{equation}
\partial _t \rho = -i(H \rho - \rho H^+) + \nd{\delta H}\rho 
\label{nqmevol}
\end{equation}
where we parametrize $\nd{\delta H}$ in a similar way to the
simple two-state system discussed earlier, namely as
\begin{equation}
  {\nd h}_{\alpha\beta} =\left( \begin{array}{c}
 0  \qquad  0 \qquad 0 \qquad 0 \\
 0  \qquad  0 \qquad 0 \qquad 0 \\
 0  \qquad  0 \qquad -2\alpha  \qquad -2\beta \\
 0  \qquad  0 \qquad -2\beta \qquad -2\gamma \end{array}\right)
\label{deltah}
\end{equation}
where the indices $\alpha, \beta$ label Pauli matrices
$\sigma_{\alpha, \beta}$ in the $K_{1,2}$ basis, and we
have assumed that $\nd{\delta H}$ has $\Delta S = 0$. It is
easy to verify that the asymptotic form (\ref{qmkl})
of the density matrix is now replaced by\footnote{We now set 
$\delta M = \delta \Gamma =0$ for simplicity: if desired, they
may be retained in a combined discussion of coherent and
decohering CPT violation.} 
\be
\rho _L
\propto \left( \begin{array}{c} 1 \qquad  \qquad
\frac{-\frac{1}{2}i  (Im \Gamma _{12} + 2\beta )- Im M_{12} }
{\frac{1}{2} \Delta \Gamma + i \Delta M } \\
\frac{\frac{1}{2}i (Im \Gamma _{12} + 2\beta )- Im M_{12} }
{\frac{1}{2}\Delta \Gamma  - i \Delta M} \qquad \qquad
|\epsilon |^2 + \frac{\gamma}{\Delta \Gamma } -
\frac{4\beta Im M_{12} (\Delta M / \Delta \Gamma ) + \beta ^2 }
{\frac{1}{4} \Delta \Gamma ^2 + \Delta M^2 } \end{array} \right)
\label{nqmkl}
\end{equation}
This is clearly a mixed state: the parameters $\alpha, \beta, \gamma$
can be regarded as causing ``regeneration in vacuo".

\pr
It is easy to see that these parameters also violate CPT~\cite{emncpt}. 
In the neutral kaon system, the CPT transformation acts as follows
on $K$ and $\bar K$ wave functions:
\begin{equation}
CPT | K > \, = \, e^{i \phi} |{\bar K}> \qquad 
CPT |{\bar K}> \, = \, e^{-i \phi} |K> 
\label{CPTK}
\end{equation}
The CPT transformation for density matrices may be represented by
\begin{equation}
   CPT  \equiv \left( \begin{array}{c}
 0   \qquad  e^{i\phi} \\
 e^{-i\phi}  \qquad 0 \end{array}\right)
\label{CPTmatrix}
\end{equation}
in the $K,{\bar K}$ basis, which is proportional to a linear
combination of $\sigma_{1,2}$. We see in this representation that
the CPT operator does not commute with $\delta m, \delta \Gamma$,
whose contributions to the Hamiltonian are proportional to
$\sigma_3$ in the $K,{\bar K}$ basis. Transforming to the
$K_{1,2}$ basis used in (\ref{deltah}) above, the CPT
operator becomes proportional to a combination of
$\sigma_{2,3}$. Since the new terms $\alpha, \beta, \gamma$
in (\ref{deltah}) couple to the indices $2,3$, they clearly violate
CPT, though in a different way from the quantum-mechanical
parameters $\delta m, \delta \Gamma$.

\pr
Experimental observables may be represented in the density
matrix formalism by expectation values of operators:
\begin{equation}
<O> \, = \, \hbox{Tr} (O \rho)
\label{expect}
\end{equation}
Common observables are represented by the operators
\bea
 &~&O_{2\pi} =\left( \begin{array}{c} 0 \qquad 0 \\
0 \qquad 1 \end{array} \right) \nn \\
&~&O_{3\pi}
 =(0.22)
 \left( \begin{array}{c} 1 \qquad 0 \\
0 \qquad 0 \end{array} \right) \nn \\
&~&O_{\pi ^-l^+ \nu} = \left(\begin{array}{c} 1 \qquad 1 \\
1\qquad 1 \end{array}\right) \nn \\
&~&O_{\pi ^+l^-{\overline \nu}}  =\left( \begin{array}{c}

1 \qquad -1 \\
 -1\qquad 1 \end{array} \right) 
\label{observe}
\eea
(strictly speaking, there should be a
corresponding prefactor of $0.998$ in the formula for
the $O_{2{\pi}}$ observable.). 
Then, for example, the semileptonic decay asymmetry
\begin{equation}
\delta \equiv \frac{\Gamma (\pi^-l^+\nu) - \Gamma (\pi^+ l^-
{\overline \nu }) }{\Gamma (\pi ^- l^+ \nu ) +
\Gamma (\pi ^+ l^- {\overline \nu})}
\label{defdelta}
\end{equation}
becomes
\bea
\nn
\delta_L &=&     2Re [\epsilon (1-\frac{i\beta}{Im M_{12}})] \\
\nn    \delta_S  &=&
2Re [ \epsilon (1 + \frac{i\beta}{Im M_{12}})]  \\
\label{semileptasy}
\eea
in the short- and long-lived kaon limits, respectively. A
difference between $\delta_{S,L}$ is often mentioned~\cite{dalitz} as a
possible signature of the CPT-violating parameter $\delta m$,
and here we see how it could also appear in our different
formalism for CPT violation.

\pr
The asymmetries which have been used so far in experimental
probes of this formalism are the $2 \pi$ decay asymmetry
\begin{equation}
A_{2 \pi} = \frac{\hbox{Tr}(O_{2 \pi}{\bar \rho}(t)) \, - \,
\hbox{Tr}(O_{2 \pi} \rho(t))}
{\hbox{Tr}(O_{2 \pi}{\bar \rho}(t)) \, + \hbox{Tr}(O_{2 \pi} \rho(t))}
\label{twopiasy}
\end{equation}
where $\rho, {\bar \rho}$ denote the density matrices of states
that are tagged initially as pure $K, {\bar K}$ respectively, 
and the double semileptonic decay asymmetry
(in an obvious short-hand notation for the rates of
different semileptonic decays) 
\begin{equation}
A_{\Delta m}={R(K^0\to\pi^+)+R(\bar K^0\to\pi^-)-R(\bar K^0\to\pi^+)
-R(K^0\to\pi^-)\over R(K^0\to\pi^+)+R(\bar K^0\to\pi^-)+R(\bar
K^0\to\pi^+)
+R(K^0\to\pi^-)}
\label{deltamasy}
\end{equation}
in which various systematic effects cancel.
As can be seen in Fig.~2, $A_{2 \pi}$ is sensitive to
the presence of $\alpha, \beta$ and $\gamma$, whereas
$A_{\Delta m}$ is particularly sensitive to $\alpha$.
The Table shows how the form of decohering CPT violation
that we propose here may be distinguished in principle
from ``conventional" quantum-mechanical CPT violation via
the parameters $\delta m, \delta \Gamma$, by measuring the
full gamut of observables discussed in \cite{ELMN,HP}.

\begin{table}[t]
\begin{center}
\caption[]{Qualitative comparison of predictions for various observables
in CPT-violating theories beyond (QMV) and within (QM) quantum mechanics.
Predictions either differ ($\ne$) or agree ($=$) with the results obtained
in conventional quantum-mechanical CP violation. Note that these frameworks can
be qualitatively distinguished via their predictions for
the asymmetries $A_{\rm T}$, $A_{\rm
CPT}$, $A_{\Delta m}$, and the interference coefficient $\zeta$
discussed in \cite{ELMN}.}
\label{Table}
\smallskip
\begin{tabular}{lcc}
\underline{Process}&QMV&QM\\
$A_{2\pi}$&$\ne$&$\ne$\\
$A_{3\pi}$&$\ne$&$\ne$\\
$A_{\rm T}$&$\ne$&$=$\\
$A_{\rm CPT}$&$=$&$\ne$\\
$A_{\Delta m}$&$\ne$&$=$\\
$\zeta$&$\ne$&$=$
\end{tabular}
\end{center}
\hrule
\end{table}

\begin{figure}
\hglue2.5cm
\epsfig{figure=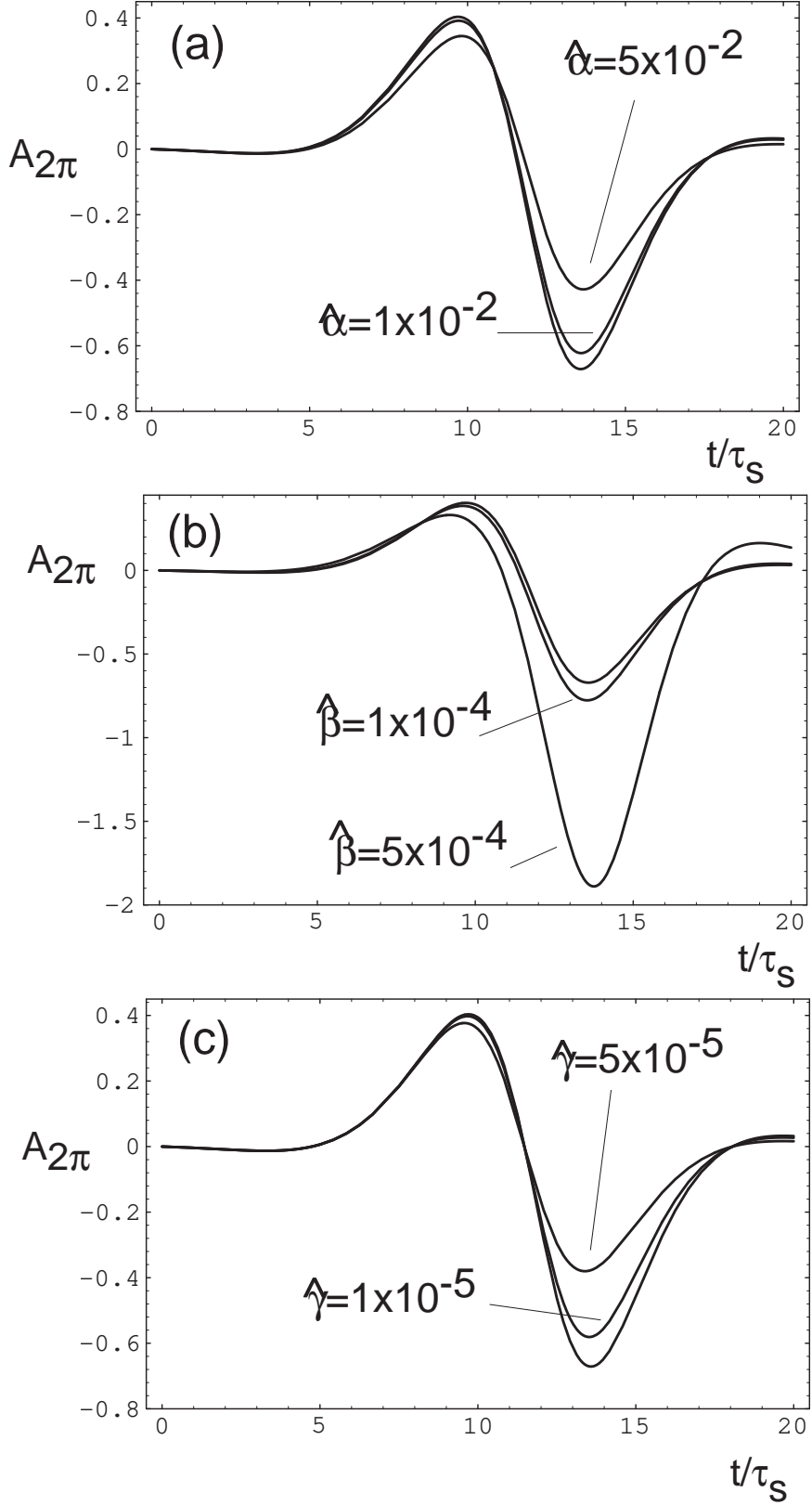,width=12cm}
\caption[]{The time-dependence of the
$2 \pi$ decay asymmetry of neutral kaons,
$A_{2 \pi}$, computed with non-zero values of
(a) $\alpha$, (b) $\beta$ and
(c) $\gamma$. We use the notation
 $\hat \alpha = \alpha/|\Delta \Gamma|$, etc., where
$\Delta \Gamma = \Gamma_S - \Gamma_L$ \cite{ELMN}.}
\end{figure}

\section{Analysis of CPLEAR Data}

\pr
Together with the CPLEAR collaboration itself, we have published 
a joint analysis of CPLEAR data~\cite{encplear}, 
constraining the CPT-violating
parameters $\alpha, \beta, \gamma$. Fig.~3 compares the data
for $A_{2 \pi}$ and $A_{\Delta m}$ compared with a
conventional quantum-mechanical fit (which is, of course,
perfectly good) and a fit in which our parameters are taken to
have values a factor of $10$ larger than the experimental limits 
we quote. Imposing the positivity constraints (\ref{positiv}),
we find the upper limits (see also the talk here by Pavlopoulos
\cite{others})

\begin{figure}
\hglue4.5cm
\epsfig{figure=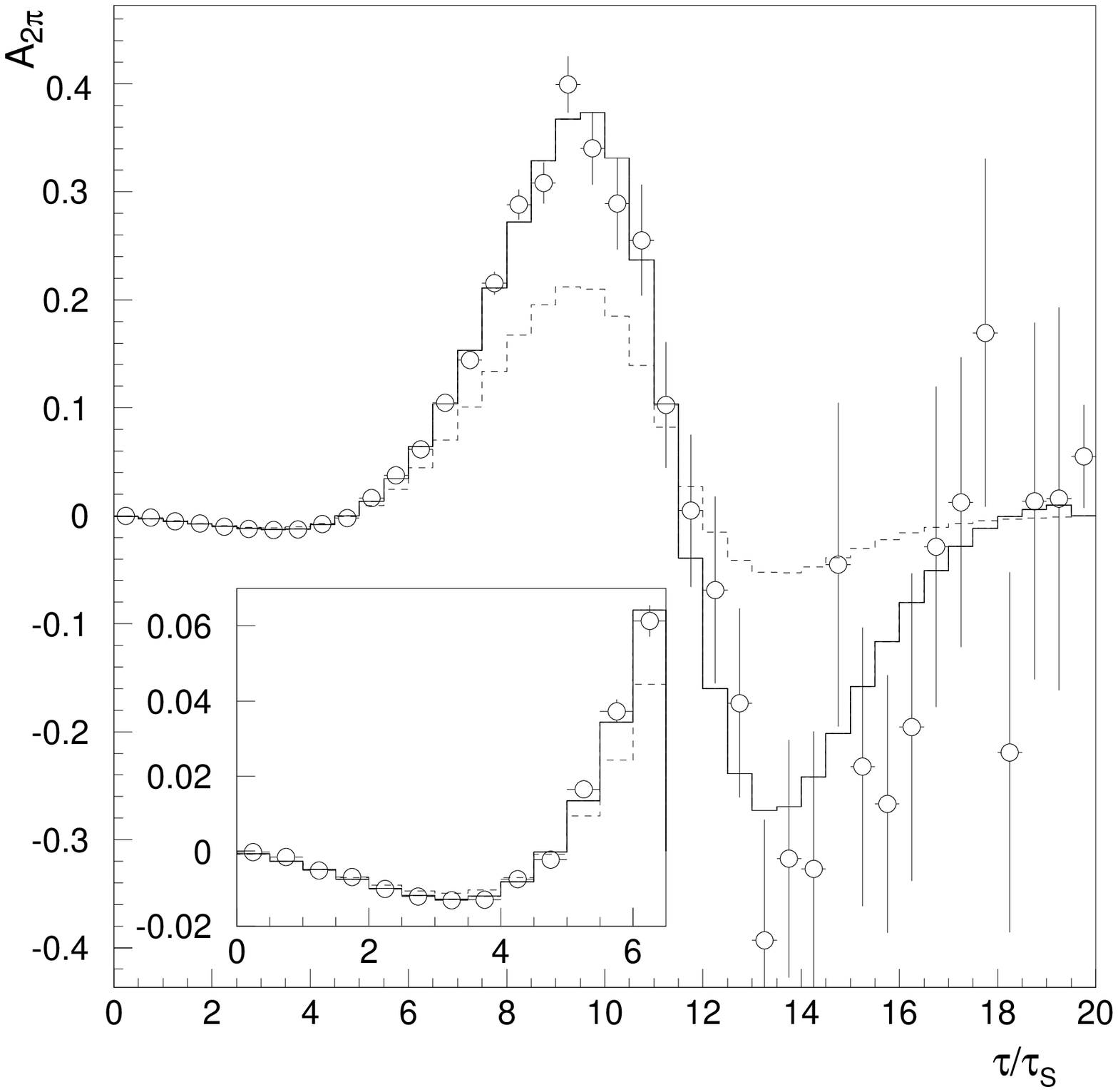,width=8cm}
\hglue4.5cm
\epsfig{figure=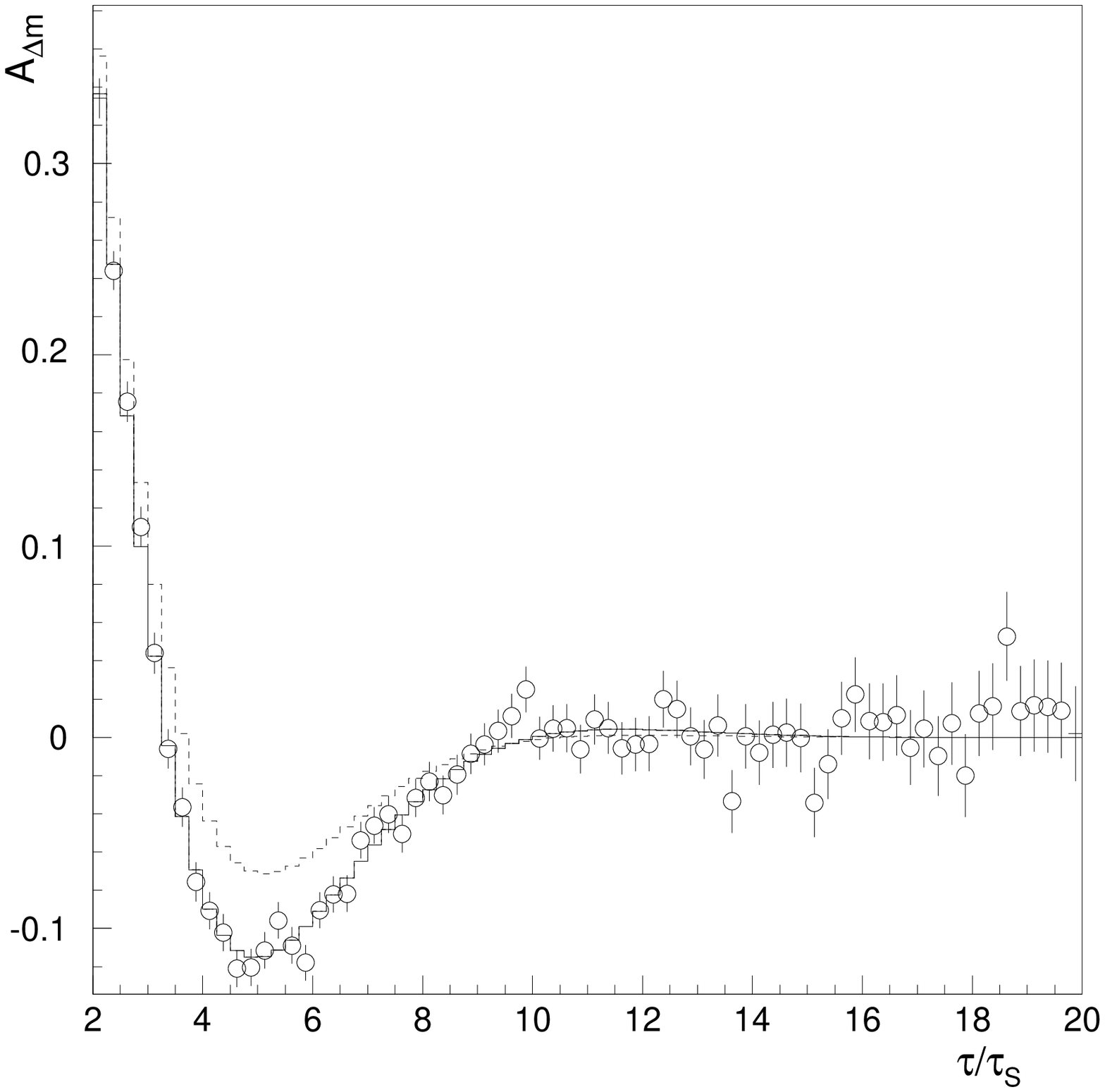,width=8cm}
\caption[]{The measured time-dependences of the neutral kaon decay
asymmetries (a) $A_{2 \pi}$ and (b) $A_{\Delta m}$, compared with
a CPT-invariant fit (solid lines) and a fit in which our bounds
on the CPT-violating parameters are
each relaxed by a factor of $10$ (dashed lines) \cite{encplear}.}
\end{figure}

\begin{equation}
\alpha < 4.0 \times 10^{-17} \hbox{GeV}, \qquad 
\beta < 2.3 \times 10^{-19} \hbox{GeV}, \qquad 
\gamma < 3.7 \times 10^{-21} \hbox{GeV}
\label{bounds}
\end{equation}
We cannot help being impressed that these bounds are in the
ballpark of $m_K^2 / M_P$, which is the maximum 
magnitude (\ref{order}) that we could expect any such effect to have.
Perhaps, with a bit more effort by CPLEAR~\cite{regen} or at 
DA$\phi$NE~\cite{HP}, ...?

\section{Outlook}

\pr
We believe there are indications from quantum gravity that the effective
quantum-mechanical description of low-energy observable laboratory
systems may need to be modified, but we remind you that this
suggestion is controversial. We have found evidence in an analysis
of non-critical string theory that supports this suggestion, but we
remind you that this is even more controversial. We welcome the
fact that the opposing views on this issue are clearly
delineated. 

\pr
Our particular approach is based on the density-matrix formalism
developed above, which we have applied specifically to the
neutral kaon system. Our approach and formalism
can in principle be distinguished from others by measuring a number of
different $K, \bar K$ decay asymmetries~\cite{ELMN}. 
Although we have presented an estimate (\ref{order})
of the largest possible magnitude that any such decohering and
CPT-violating effects might have, we are not in a position to calculate
its magnitude. Even if this order of magnitude is attained for some
other microscopic system, it is conceivable that it might be suppressed
in the neutral kaon system for some `accidental' reason associated
with the way the $K$ is made out of light string modes. And, of
course, we cannot exclude the possibility that
the decohering and CPT-violating effects we discuss may be
suppressed by additional powers of $m_K/M_P$ (or even an exponential!)
below our maximal estimate (\ref{order}). Therefore we can offer
our experimental colleagues no guarantee of success. Nevertheless,
we think that the importance of the issues
discussed here motivate a new series of microscopic experiments to
test quantum mechanics and CPT. We are glad that other speakers   
at this meeting agree with us, even if their approaches to the
issues differ from ours.

\pr
{\bf Acknowledgements}

\pr
One of us (J.E.) thanks the organizers for their invitation to
speak at this meeting, and thanks the Lawrence Berkeley National
Laboratory and the Berkeley Center for Particle Astrophysics for
their hospitality while it was being written. His work is 
supported in part by the Director, Office of Energy Research,
Office of Basic Energy Science of the U.S. Department of Energy,
under Contract DE-AC03-76SF00098.


\begin{thebibliography}{99}
\bibitem{lud} G. L\"uders, Ann. Phys. (N.Y.) 2, (1957), 1;
\par W. Pauli, {\it Niels Bohr and the Development 
of Physics}, eds. W. Pauli, L. Rosenfeld, and V. Weisskopf
(Mc. Graw Hill, New York (1955));
\par R. Jost, Helv. Phys. Acta 31 (1958), 263. 
\bibitem{pdg} {\it Review of Particle Properties},
Particle Data Group, Phys. Rev. D54 (No 1, part I) (1996), 1.
\bibitem{dmk} CPLEAR Collaboration, R. Adler {\it et al.},
{\it An improved determination of the $K^0$-${\overline K}^0$
mass difference: a test of $CPT$ symmetry}, in preparation.  
\bibitem{hawkbek} S. Hawking, Comm. Math. Phys. 43 (1975), 199;
\par J. Bekenstein, Phys. Rev. D12 (1975), 3077.
\bibitem{gross} D.J. Gross, Nucl. Phys. B236 (1984), 349.
\bibitem{hawk2} S. Hawking, Comm. Math. Phys. 87 (1982), 395.
\bibitem{ehns} J. Ellis, J.S. Hagelin, D.V. Nanopoulos and
M. Srednicki, Nucl. Phys. B241 (1984), 381.
\bibitem{BPS} T. Banks, M.E. Peskin and L. Susskind,
Nucl. Phys. B244 (1984), 125.
\bibitem{emn} J. Ellis, N. E. Mavromatos and 
D.V. Nanopoulos, Phys. Lett. B293 (1992), 37;
\par {\it Lectures presented at the Erice Summer 
School, 31st Course: From Supersymmetry 
to the Origin of Space Time}, Ettore-Majorana
Centre, Erice (Italy), July 4-12 1993, published 
in Proc. Subnuclear Series Vol. 31, p.1 (World-Sci.,
Singapore 1994);
\par Mod. Phys. Lett. A10 (1995), 425; and hep-th/9305117. 
\bibitem{unruh} W. Unruh and R. Wald, Phys. Rev. D52 (1995), 2176.
\bibitem{wald} R. Wald, Phys. Rev. D21 (1980), 2742;
\par D. Page, Gen. Rel. Gravit. 14 (1982), 299. 
\bibitem{hpbook} S. Hawking and R. Penrose, {\it The Nature of 
Space and Time} (Princeton University Press, 1996).
\bibitem{wittson} E. Witten, Comm. Math. Phys. 109 (1987), 525;
\par H. Sonoda, Nucl. Phys. B326 (1989), 135.
\bibitem{kostel} V.A. Kosteleck\'y and R. Potting, 
Nucl. Phys. B359 (1991), 545. 
\bibitem{others} See the contributions to these proceedings by
P. Huet, V.A. Kostaleck\'y, L. Okun, N. Pavlopoulos and Y. Tsai.
\bibitem{polch} See, for instance: J. Polchinski, 
{\it String Duality - A Colloquium}, preprint NSF-ITP-96-60,
hep-th/9607050;
J.H. Schwarz, {\it The Second Superstring Revolution}, hep-th/9607067, 
and references therein. 
\bibitem{emnstates} J. Ellis, N. E. Mavromatos 
and D.V. Nanopoulos, Phys. Lett. B278 (1992), 246. 
\bibitem{emnmeas} J. Ellis, N. E. Mavromatos 
and D.V. Nanopoulos, Phys. Lett. B284 (1992), 27; 
{\it ibid.} 43. 
\bibitem{banksd} T. Banks, Rutgers Univ. preprint RU-96-49,
hep-th/9606026. 
\bibitem{chaulykk} S. Chaudhuri and J. Lykken, 
Nucl. Phys. B396 (1993), 270. 
\bibitem{zam} A.B. Zamolodchikov, JETP Lett. 43 (1986), 730;
Sov. J. Nucl. Phys. 46 (1987), 1090.
\bibitem{kogan} I. Kogan, preprint UBCTP-91-13 (1991).
\bibitem{mohanty} J. Ellis, S. Mohanty and D.V. Nanopoulos, 
Phys. Lett. B221 (1989), 113; 
\par {\it ibid.} B235 (1990), 305. 
\bibitem{amelino} G. Amelino-Camelia, 
J. Ellis, N.E. Mavromatos and D.V. Nanopoulos, 
preprint ACT-07/96, CERN-TH/96-143,CTP-TAMU-21/96,OUTP-96-24P,
hep-th/9605211. This article also points out that the space-time
foam effects discussed here are compatible with known aspects of
distance measurements.
\bibitem{elizabeth} J. Ellis, N.E. Mavromatos, D.V. Nanopoulos
and E. Winstanley,  
preprint ACT-02/95, CERN-TH/95-336,CTP-TAMU-04/95,OUTP-96-05P,
gr-qc/9602011. 
\bibitem{dbrane} J. Ellis, N.E. Mavromatos and D.V. Nanopoulos, 
preprint ACT-04/96, CERN-TH/96-81,CTP-TAMU-11/96,OUTP-96-15P,
hep-th/9605046, Int. J. Mod. Phys. A in press. 
\bibitem{dalitz} N.W. Tanner and R.H. Dalitz, 
Ann. Phys. (N.Y.) 171 (1986), 463;
\par C.D. Buchanan, R. Cousins, C. O. Dib,
R.D. Peccei and J. Quackenbush, Phys. Rev. D45 (1992), 4088;
\par C.O. Dib and R.D. Peccei, Phys. Rev. D46 (1992), 2265.
\bibitem{emncpt} J. Ellis, N.E. Mavromatos and D.V. Nanopoulos,
Phys. Lett. B293 (1992), 142;
\par Int. J. Mod. Phys. A11 (1996), 1489. 
\bibitem{ELMN} J. Ellis, J. Lopez, N.E. Mavromatos and 
D.V. Nanopoulos, Phys. Rev. D53 (1996), 3846. 
\bibitem{HP} P. Huet and M.E. Peskin, Nucl. Phys. B434 (1995) 3.
\bibitem{regen} J. Ellis, J. Lopez, N.E. Mavromatos and D.V. Nanopoulos,
CERN preprint CERN-TH/96-82 (1996).
\bibitem{encplear} CPLEAR Collaboration, R. Adler {\it et al.},  
and J. Ellis, 
J. Lopez, N.E. Mavromatos and D.V. Nanopoulos, 
Phys. Lett. B364 (1995), 239. 
\end{thebibliography}
\end{document}